\documentclass[10pt,conference]{IEEEtran}
\usepackage[T1]{fontenc}
\usepackage[utf8]{inputenc}
\usepackage{xurl}
\usepackage{amsmath,amssymb,booktabs,array,ragged2e,microtype,graphicx,tikz}
\usetikzlibrary{arrows.meta,positioning,calc}
\usepackage{hyperref}
\newcommand{\code}[1]{\texttt{#1}}
\begin{document}
\title{Does the Judge See What the Evaluation Requires? Input Assurance for LLM Agent Evaluation}

\author{\IEEEauthorblockN{Boming Xia\IEEEauthorrefmark{1},
Qinghua Lu\IEEEauthorrefmark{2},
Runze Li\IEEEauthorrefmark{3},
Dino Sejdinovic\IEEEauthorrefmark{4} and
Liming Zhu\IEEEauthorrefmark{2}}
\IEEEauthorblockA{\IEEEauthorrefmark{1}Responsible AI Research Centre, Adelaide University, Australia\\
\IEEEauthorrefmark{2}CSIRO, Australia\\
\IEEEauthorrefmark{3}UNSW Sydney, Australia\\
\IEEEauthorrefmark{4}Adelaide University, Australia\\
}
}

\maketitle
\raggedbottom

\begin{abstract}
An LLM judge assesses agent steps from text packets. A renderer converts recorded runs into these packets, while software checkers test records and packets against requirements. We ask what evidence shows a packet is what the evaluation requires and propose four input-assurance obligations: required content survives rendering, checkers cover their requirements, a failed check stops later use and the judge reads the bytes last checked. In one pipeline with 30 trajectories, matching hashes, passing checkers and successful staging (copying packets for the judge) coexisted with violations of all four obligations, the last two in integration tests. The renderer omitted every structured tool call, leaving 53 of 60 judged steps blank despite matching hashes. For one judge configuration, later context raised error flags by 13 more on error than on correct steps. Restoring every call reduced this contrast to 2, a reduction of 11 (0.37 per trajectory, 95\% cluster-bootstrap interval 0.17 to 0.53). In packets ending at the judged step, that configuration abstained at least once on every blank step, while the other four sometimes abstained but also flagged blank error and correct steps. Both separately written checkers accepted an unmatched tool result. In integration tests, a stub replacing the judge read a failed build's packet and a copy changed after staging. Three added controls stopped 18 of 19 constructed fault cases before the stub read any packet. The remaining stale script never ran. We turn the obligations into four tests and plan to evaluate their transfer to independent pipelines.
\end{abstract}


\section{Introduction}\label{sec:intro}

LLM judges are used to assess individual steps in recorded runs (trajectories) of tool-using agents~\cite{lu2025agentrewardbench,apb2026}.
In the pipeline we study, a renderer turns structured trajectory records into text packets for the judge.
Two software checkers test source records and packets against written requirements.
An agent may leave its message text empty and place its action only in a structured tool-call field.
If the renderer copies only message text, it omits the call and makes the judged step appear blank.
The packet can still match its recorded hash, because a hash shows only that the bytes are unchanged.

We ask what evidence establishes that a judge's input meets the evaluation's requirements, and call such evidence \emph{input assurance}.
The content and validation rules that this input must satisfy form the judge's \emph{input contract}.
We propose four obligations for the software that prepares that input: (1)~required content survives rendering, (2)~each checker covers its stated requirements, (3)~a failed check stops later use and (4)~the judge reads the bytes last checked.
Each requires distinct evidence: comparison with source records for content, violating cases for requirements coverage, integration tests for failure handling and byte checks at use for integrity.

The pipeline compares judgements on trajectory prefixes (ending at the judged step) with judgements on full trajectories, to assess how later context changes error flags.
Its renderer kept tool results but omitted every structured tool call, leaving 53 of 60 judged steps blank in the original packets.
We compared these with call-restored packets, which restore calls throughout each packet, under five judge configurations (model, prompt and delivery settings).
With the original packets, moving from prefix to full packets raised error flags under the first configuration, J0, by 15 on error steps and by 2 on correct steps, a contrast of 13.
With the calls restored, this contrast was 2.

Research software can produce reproducible results despite bugs~\cite{papi2024giant}, and direct data validation can expose problems not apparent from model quality~\cite{breck2019data}.
Work on evaluation harnesses addresses implementation choices and reproducibility~\cite{biderman2024lessons}, while benchmark audits identify flaws in task setup and grading~\cite{zhu2025rigorous}.
AgentRewardBench shows that trajectory representation affects judge agreement with experts~\cite{lu2025agentrewardbench}.
Judge-robustness studies examine deliberately changed inputs~\cite{dev2026judge}, and content-free responses can draw favourable verdicts~\cite{zheng2025cheating,zhao2025onetoken}.
Fault-injection studies examine agent failures~\cite{tan2026agentchaos,seyedghorban2026observability}, while SPADE synthesises and selects assertions for LLM pipeline outputs~\cite{shankar2024spade}.
Unlike these studies, we examine whether the judge's input contract is preserved and enforced from rendering to use, and what the pipeline's own checks establish about the bytes the judge reads.
We assess the two separately written checkers as test oracles~\cite{jahangirova2016test} and use a recording stub in place of the judge to observe the bytes read in integration tests.

On the real packets both checkers passed, and staging (copying packets for the judge and checking their hashes) succeeded.
Yet mutation cases showed that both accepted a tool result with no matching call, and code inspection found a hash comparison that could never fail.
Constructed integration tests showed that failed or changed packets could still reach the point of use.
These defects reflect established software-engineering concerns: incomplete test oracles~\cite{barr2015oracle}, shared failures in separately written programs~\cite{knight1986experimental}, redundant assertions~\cite{peruma2019distribution}, build failure handling~\cite{gnumake2023} and time-of-check-to-time-of-use gaps~\cite{dean2004races}.
Output-based testing requires a fault's effects to reach an observed result~\cite{voas1992pie}, but the judges' answers, although affected by the omission, did not identify the missing calls.
In prefix packets, whether a blank step drew abstentions or error flags depended on the judge configuration.

The paper makes three contributions.
It organises established assurance practices into four obligations for the boundary between recorded trajectories and judge inputs, with a direct test for each (Table~\ref{tab:diagnosis}).
It shows a rendering omission preserved by matching hashes, requirements missed by passing checkers and failed or changed packets read in integration tests, and measures how restoring the calls changed a judge-based contrast.
It evaluates 3 controls on 19 constructed fault cases and proposes an evaluation of the tests' transfer to independent pipelines.

\section{Case and Study Design}\label{sec:design}
\subsection{Targets and renderings}

Each judgement concerns one agent step, the \emph{target step}, marked in a text \emph{packet} containing task context and recorded trajectory messages.
A \emph{prefix packet} ends at that step, before its tool result, while a \emph{full packet} continues to the trajectory's end.
We use the BFCL (Berkeley Function-Calling Leaderboard)~\cite{patil2025bfcl} and tau2~\cite{barres2025tau2} (customer service with simulated-user tool calls) subsets of AgentProcessBench~\cite{apb2026}.
The benchmark labels agent steps $+1$ (correct and effective), $0$ (neutral) or $-1$ (incorrect or harmful).
Steps depending on an error remain $-1$ until it is corrected or an independent subtask begins.
The labels select targets but are not included in judge packets.

\begin{figure*}
\centering
\begin{tikzpicture}[
 stage/.style={draw,rounded corners=1pt,minimum height=6mm,text width=18mm,align=center,font=\footnotesize,inner sep=2pt},
 check/.style={draw,dashed,fill=gray!7,align=center,font=\footnotesize,inner sep=3pt,text width=34mm},
 arr/.style={-{Latex[length=2mm]},semithick}]
\node[stage] (src) {Structured\\trajectory};
\node[stage,right=3.5mm of src] (build) {Build R\\packets};
\node[stage,right=3.5mm of build] (arc) {O/R\\archive};
\node[stage,right=3.5mm of arc] (stage) {Staging:\\copy + hash};
\node[stage,right=3.5mm of stage] (copies) {Staged\\packet files};
\node[stage,right=7mm of copies,text width=25mm] (use) {Dispatch:\\judge reads copy};
\foreach \a/\b in {src/build,build/arc,arc/stage,stage/copies,copies/use} \draw[arr] (\a) -- (\b);
\node[check,above=5mm of build,text width=48mm] (inline) {Inline checker:\\source, in-memory packets, O archive};
\node[check,above=5mm of stage] (disk) {Disk checker:\\source + archived files};
\draw[arr,dashed] (build.north) -- (inline.south);
\draw[arr,dashed] (arc.north) -- ++(0,3mm) -| (disk.south);
\node[stage,below=2mm of arc,minimum height=4.5mm,text width=31mm] (man) {Manifest + job list};
\node[stage,at={(copies |- man)},minimum height=5mm,text width=22mm] (emit) {Emit scripts};
\draw[arr] (build.south) |- (man.west);
\draw[arr] (man.east) -- (emit.west);
\draw[arr] (stage |- man) -- (stage.south);
\fill (stage |- man) circle (0.5pt);
\draw[arr] (emit.east) -| (use.south);
\end{tikzpicture}
\par
\begin{minipage}[t]{0.30\textwidth}
\footnotesize\textbf{Source: selected fields}
\begin{verbatim}
"content": "",
"tool_calls": [{"function": {
  "name": "cd", "arguments":
  "{\"folder\":\"..\"}"}}]
\end{verbatim}
\end{minipage}\hfill
\begin{minipage}[t]{0.28\textwidth}
\footnotesize\textbf{O: original packet}
\begin{verbatim}
[BEGIN TARGET STEP]
[assistant]

[END TARGET STEP]
\end{verbatim}
\end{minipage}\hfill
\begin{minipage}[t]{0.36\textwidth}
\footnotesize\textbf{R: calls restored}
\begin{verbatim}
[BEGIN TARGET STEP]
[assistant]
[TOOL CALL] cd({"folder":".."})
[END TARGET STEP]
\end{verbatim}
\end{minipage}
\caption{Pipeline before the added controls (top) and one error target, \code{bfcl:527}, in its source record and prefix packets (bottom). The original (O) packet omits the target's tool call, and the call-restored (R) packet includes it. In both draws, J0 abstained on O and answered \code{yes} on R. Dashed arrows locate the checkers, and their boxes list the inputs they read. Downstream steps did not enforce checker outcomes.}
\label{fig:pipeline}
\end{figure*}

Each selected trajectory contributes two target steps.
The \emph{error target} is the earliest tool-calling step that begins a run of $-1$ labels.
The \emph{comparison target} is the closest earlier tool-calling step labelled $+1$.
Of 500 trajectories, 99 remained after applying these rules and excluding 32 used in earlier runs.
An earlier study sampled 30 of them at random with a fixed seed, 14 BFCL and 16 tau2, and we reuse that cohort (supplementary Table~S1 and Section~I).

The original renderer kept message roles and text but omitted the structured tool-call field (\code{messages[].tool\_calls}).
A target with empty message text therefore appeared blank: 26 of 30 error targets and 27 of 30 comparison targets (53 of 60).
The other seven had text but still lost their calls.

We compared original (O) and call-restored (R) packets (Fig.~\ref{fig:pipeline}).
O reuses the earlier study's archived packets byte for byte, and each still matched its recorded hash.
R inserts each missing call as \mbox{\code{[TOOL CALL]}} \code{name(arguments)} after the text of its originating message, preserving call order and arguments.
R restores calls throughout each packet, not only at the target, so the O/R comparison concerns the whole rendering.
In both renderings, tool results omit their tool names and the ids linking them to calls, and tau2 packets omit tool definitions.
R therefore restores call lines but is not a complete rendering (supplementary Section~I).
The two target types, two packet extents and two renderings give eight cells of 30 packets, or 240 packets in total.
Each judge configuration scheduled two separate judgements (\emph{draws}) per packet, or 480 per configuration, and a consensus error flag required both draws to answer \code{yes}.
For each rendering, the contrast is $G=(F_E-P_E)-(F_C-P_C)$, where $F$ and $P$ count consensus flags in full and prefix packets, and $E$ and $C$ denote error and comparison targets.

\subsection{Pipeline and evidence}

The builder creates the R packets and writes a \emph{manifest} of expected hashes for all 240 O/R packets and a \emph{job list} assigning each draw its packet.
Staging copies the packets into a separate folder and checks hashes of their text against the manifest.
Emission turns the job list into scripts identifying each draw's packet, and dispatch runs them.
Two software checkers validate source records and packets, rather than judge agent behaviour.
The \emph{inline checker} runs inside the builder, before staging, while the \emph{disk checker} later rereads source records and archived O/R packets.
This study's written plan assigned both checkers ten requirements, including call placement and staged-file hashes (supplementary Section~II).
The two checkers were written from the same plan but separately, and the disk checker's writer did not read the builder's code.

Both checkers reported that all ten requirements were met on the real packets, and J0 judged all 240 packets in the diagnostic run (Section~\ref{sec:diag}).
We read both checkers' code and ran them on valid inputs and on 38 constructed \emph{mutation cases}, each intended to violate a requirement by changing source records, in-memory packets or saved files.
We counted a miss when a checker reported no problem while the mutation was still present.
The \emph{shared mutation cases} are the 34 cases whose changes reached inputs read by both checkers (supplementary Section~II).

Twelve \emph{integration tests} examined whether packets could still reach later steps after a failed check or file change.
We separately ran 19 constructed fault cases under four cumulative conditions: P0 without added controls and P1 to P3 with controls added successively (Section~\ref{sec:enf}).
Eight cases reused mutation cases or integration tests, and eleven used new injection points chosen by a fixed rule.
A \emph{recording stub} replaced the judge in these tests and hashed any packet it read, showing which bytes reached the point of use.

\begin{table*}
\centering\footnotesize
\caption{Obligations for input assurance with their tests (Section~\ref{sec:disc}), what the case showed and the controls. Row 1 uses the real packets, and rows 2 to 4 code inspection and constructed cases. Tested: exercised in this study. Recommended: not exercised.}
\label{tab:diagnosis}
\begin{tabular}{@{}>{\raggedright\arraybackslash}p{0.32\textwidth}>{\raggedright\arraybackslash}p{0.36\textwidth}>{\raggedright\arraybackslash}p{0.27\textwidth}@{}}
\toprule Obligation and test & What the case showed & Control and status\\\midrule
1. Required content survives rendering. Test: render an action held only in a structured field & Packets omitted every call yet matched their hashes & Restored call lines (R): \textbf{tested}\\
2. Checkers cover their requirements. Test: violate each requirement on the object and at the pipeline step it covers & Both accepted an injected unmatched result, and neither read the staged copies & Reverse matching (P3): \textbf{tested} on three cases. Disk-checker fixes: \textbf{recommended}\\
3. A failed check stops later use. Test: continue after a failure & A failed build's packet was staged and read & Stop on failure (P1): \textbf{tested}\\
4. The judge reads the bytes last checked. Test: change a checked file & The stub read a copy changed after staging and a packet with line endings changed before staging & Checks before use (P2): \textbf{tested} after staging. Build-time exact-byte hashes checked at staging: \textbf{recommended}\\
\bottomrule
\end{tabular}
\end{table*}

\section{What Passing Checkers Did Not Show}\label{sec:conf}

A checker may pass without covering every assigned requirement (obligation~2).
Coverage depends on which properties it checks, which files it reads and when it runs.
Both checkers accepted the real packets, but code inspection and mutation cases showed that neither checked for tool results without matching calls or inspected the staged copies.

\subsection{A requirement checked in only one direction}

One requirement called for one-to-one matching between tool calls and results in the source records.
Both checkers verified that each call had a result, but not that each result had a call.
The inline checker's loop starts from calls and skips messages without them:
\par\smallskip\noindent\begin{minipage}{\columnwidth}\small
\begin{verbatim}
tcs = m.get("tool_calls") or []
if not tcs:
    continue
\end{verbatim}
\end{minipage}\par\smallskip
When we added a tool result with no matching call to a source record and rebuilt the packets, both checkers exited successfully.
This was the only one of the 34 shared mutation cases that both accepted.
Across the 34 cases, the inline checker exited successfully despite two violations and the disk checker despite one.
Their error output pointed to the violated requirement in 23 and 30 cases, respectively, and their other nonzero exits named other failures or crashed (supplementary Table~S3).
The disk checker rejected an invalid JSON argument string that the inline checker accepted.
The shared omission illustrates why separately written checkers based on one plan cannot be assumed to fail independently~\cite{knight1986experimental}.

\subsection{A requirement checked on the wrong files}

The plan required 240 staged files, each with a hash matching its manifest entry, and assigned this requirement to both checkers.
The inline checker runs before staging creates those files, so it cannot inspect them.
The disk checker runs later but reads archived packets, not the staged copies or the written manifest.
Only staging's own hash check compared staged files with the manifest.

The disk checker's own hash comparison built its expected hash from archive text it had loaded, then hashed the same unchanged text again.
Here \code{O} and \code{R} hold that loaded text, and \code{m} is an expected-hash entry built from it:
\par\smallskip\noindent\begin{minipage}{\columnwidth}\small
\begin{verbatim}
"sha256": sha256_hex(txt),
...
txt = (O if rend == "O" else R)[(rid, cond)]
if sha256_hex(txt) != m["sha256"]:
    sha_bad += 1
\end{verbatim}
\end{minipage}\par\smallskip
Since both hashes come from the same unchanged text, this comparison cannot detect a changed file.
It has the same failure mechanism as a redundant assertion in software tests~\cite{peruma2019distribution}.
To cover its assigned staged-file requirement, the disk checker must inspect the staged file set and compare each copy's hash with its previously recorded manifest entry (Table~\ref{tab:diagnosis}).

\begin{table*}[t]
\centering\footnotesize
\caption{Outcomes of 19 constructed fault cases under cumulative controls, with reads by the recording stub. P0 continues after a failure, P1 stops on failure, P2 also checks files before use and P3 also matches every result to a call.}
\label{tab:enforcement}
\setlength{\tabcolsep}{4pt}
\begin{tabular}{@{}>{\raggedright\arraybackslash}p{0.36\textwidth}r>{\raggedright\arraybackslash}p{0.30\textwidth}>{\raggedright\arraybackslash}p{0.25\textwidth}@{}}
\toprule
Fault and injection point & Cases & Outcome under P0 & First stop or refusal\\\midrule
Restored call deleted or changed during build & 5 & Altered packet read & P1: build\\
Restored call deleted after build & 1 & Altered packet read & P1: staging\\
Invalid argument JSON in a source record & 1 & Altered packet read & P1: disk check\\
Unmatched result added to a source record & 3 & Altered packet read & P3: reverse matching\\
Manifest hash corrupted after build & 3 & Unaltered packet read after staging failed & P1: staging\\
Failed build with previous outputs present & 1 & Previous build's output read & P1: build\\
Packet changed after staging & 3 & Altered packet read & P2: before dispatch\\
Extra file present before staging & 1 & Run completed with an extra staged file & P2: before dispatch\\
Extra stale script & 1 & Script unused & None: unused under P0--P3\\\midrule
Total & 19 & & \\\bottomrule
\end{tabular}
\par\smallskip
\begin{minipage}{0.97\textwidth}\footnotesize
Where a first stop or refusal is listed, earlier conditions retain the P0 outcome and later conditions retain that stop or refusal.
\end{minipage}
\end{table*}

\section{What Successful Staging Did Not Stop}\label{sec:enf}

Before the added controls, downstream steps did not require earlier steps to have succeeded, and nothing checked a staged packet's bytes when the judge read it (obligations~3 and~4).

\subsection{Later steps read failed and changed packets}

In an integration test, the builder wrote a packet missing the target's tool call and exited with an error.
The test deliberately continued through staging and emission.
Both succeeded, and the recording stub read the altered packet in both scheduled draws.
Staging passed because the failed build had written the packet and its manifest hash from the same altered text.
In another test, we changed a staged packet after staging passed, and the stub read the changed bytes: a time-of-check-to-time-of-use gap~\cite{dean2004races}.
A third test changed a packet's line endings to CRLF after build but before staging.
Staging hashed the packet as text, normalising CRLF to LF, so its check passed although the stub read changed bytes (supplementary Section~III).

\subsection{Different controls stop different cases}

We ran 19 constructed fault cases under four cumulative conditions (Table~\ref{tab:enforcement}).
P0 invokes later steps despite earlier failures and dispatches only scripts emitted for that run.
P1 stops when a step exits with an error or a required checker does not report success.
P2 records each staged packet's exact-byte hash at staging and checks the archived packets against those hashes before and after the disk check.
Before dispatch, it requires successful validation for this run and the expected, unchanged staged files, manifest, job list and scripts.
The recording stub also checks each packet's hash at read time (supplementary Section~IV).
P3 adds reverse matching: every tool result in the source records must answer a call, including calls by simulated users.

P1 stopped 11 cases.
P2 refused dispatch in four more: three packets changed after staging and one extra staged file.
The three unmatched-result cases passed P2 because neither checker enforced reverse matching and their files remained unchanged, and P3 stopped them.
The remaining case, an extra stale script, was unused under every condition.
Under P0, 14 of the 18 cases stopped by P1 to P3 had led to altered or stale packet reads.
Both cases without injected faults (the unaltered cohort and one with an unrelated file outside the staged folder) completed under all conditions, and every packet the stub read matched its hash in the diagnostic run's manifest.
In a separate test, the stub under P2 and P3 refused a packet changed after the last check before dispatch, in both draws.
The line-ending change made before staging was not among the 19 cases.
By code inspection, P2 and P3 would accept it: staging hashes normalised text, and exact-byte packet hashes start only at staging.

\section{What Judge Responses Did Not Show}\label{sec:diag}

Restoring calls changed J0's contrast, but the judges' responses did not identify the missing calls (obligation~1, Table~\ref{tab:main}).
Each judge was instructed to use only the supplied packet to decide whether the target step was incorrect or counterproductive for the user's task.
The permitted answers were \code{yes}, \code{no} and \code{cannot\_decide}.

\begin{table}
\centering\footnotesize
\caption{Consensus error flags (both draws \code{yes}) among 30 targets per cell, original (O) $\to$ call-restored (R) packets.}
\label{tab:main}
\setlength{\tabcolsep}{3pt}
\begin{tabular}{@{}llrrrr@{}}
\toprule
 & & \multicolumn{2}{c}{Error target} & \multicolumn{2}{c}{Comparison ($+1$)}\\
\cmidrule(lr){3-4}\cmidrule(lr){5-6}
Key & Model alias & Prefix & Full & Prefix & Full\\\midrule
J0 & \code{claude-opus-5} & $2\to20$ & $17\to22$ & $0\to2$ & $2\to2$\\
J1 & \code{gpt-5.6-terra} & $22\to19$ & $13\to18$ & $15\to9$ & $6\to3$\\
J2 & \code{gpt-5.6-luna} & $22\to20$ & $12\to21$ & $23\to7$ & $8\to5$\\
J3 & \code{claude-sonnet-5} & $8\to21$ & $9\to15$ & $7\to9$ & $4\to3$\\
J4 & \code{qwen3.6:27b} & $21\to19$ & $18\to16$ & $17\to7$ & $15\to4$\\
\bottomrule
\end{tabular}
\par\smallskip
\begin{minipage}{\columnwidth}\footnotesize
Each arrow compares cell totals (per-target changes: supplementary Table~S9).
\end{minipage}
\end{table}

J0's O and R draws ran in a randomly interleaved order to limit time-related confounding, each in a fresh session that read its packet from a file.
The recorded reads returned every packet line.
J1 to J4, run later to examine variation across configurations, received packet text in their prompts without tools.
For each J1 to J4 request, the recorded hash matched the hash computed from the manifest-listed packet bytes followed by one separating line break.
We interpret differences at the configuration level because model, prompt wrapper, output-schema enforcement, reasoning settings, client and serving software, packet delivery and date changed together (supplementary Sections~V and~VI).

For J0, the error-prefix flag count rose from 2 with O to 20 with R: 18 targets gained a consensus flag and none lost one (supplementary Table~S9).
With O packets, full packets had 15 more flags than prefix packets on error targets and 2 more on comparison targets, giving $G=13$.
Fourteen of the 17 error-full flags concerned targets blank in their original prefix packets, where J0 had abstained.
With R packets, $G=2$, a reduction of 11 (0.37 per trajectory).
Resampling the 22 clusters formed by shared user requests, with recorded verdicts fixed, gave 95\% bootstrap intervals (per trajectory) of 0.17 to 0.53 for the reduction and $-0.05$ to 0.21 for the restored contrast (supplementary Table~S7).
With O packets, J1 to J4 had contrasts of 0, 5, 4 and $-1$, compared with J0's 13.

In J0's original prefix packets, both draws abstained on all 26 blank error targets, and at least one draw abstained on each of the 27 blank comparison targets.
Such abstentions could not produce consensus error flags.
No J0 draw abstained on these targets in R prefix packets.
On the same blank O prefix targets, J1 to J4 flagged 20, 20, 7 and 19 of the 26 error targets, and 13, 22, 7 and 17 of the 27 comparison targets (supplementary Table~S8).
A consensus error flag does not reveal whether the target action was present in the packet, and an abstention does not identify the missing call.
At least one draw abstained on 17 of 30 original error-prefix packets under J3 and 3 to 6 under J1, J2 and J4, all at blank targets (supplementary Section~VI).
The benchmark labels were assigned using complete trajectories, so a flag on a blank prefix error target does not establish that the judge detected the omitted action's error.
Direct validation of packet content against its source can identify the missing calls without relying on judge responses~\cite{breck2019data}.

\section{Implications}\label{sec:disc}

An LLM judge, whether assessing agent steps or annotating software artifacts~\cite{ahmed2025annotation}, is one part of a measurement process: source selection, rendering, checking and response aggregation also shape its results.
In this case, the answer schema had no category for defective input, and J0's abstentions on blank original prefix targets counted as no error flags.
Full packets then produced flags on some of those targets, contributing to the measured contrast even though their tool calls remained omitted.
The engineering implication for construct validity~\cite{ralph2018construct} is that evaluations should report evidence that judge inputs met their requirements.
The four tests below provide direct evidence for the input contract without live judge calls, using a recording stub where needed, and can serve as regression tests when input representations or pipeline components change.
Each test includes a valid input that must stay accepted.

\textbf{Test 1: render required content.}
Derive required and excluded fields from the input contract for each packet type.
Render a message with empty text and an action only in a structured field, then check the action's name, arguments and position against the source.
Check that excluded information, such as benchmark labels or the target's tool result in a prefix packet, remains absent.
For screenshots or accessibility trees, check contract-required information against the source, allowing permitted transformations.
As secondary diagnostics, report abstentions separately from \code{no} and inspect error flags on steps labelled correct.

\textbf{Test 2: violate each requirement.}
For each requirement, specify its object, pipeline step and expected behaviour, then test its predicates with cases expected to pass and fail.
For one-to-one relations, test both directions: a call without a result and a result without a call.
A content check should take expected values from source records, while an integrity check should compare the current bytes' hash with a reference recorded before any change it must detect.
The contract must match the evaluation goal: valid JSON was required here, but malformed calls should be preserved when their behaviour is being judged.
Requirements and checks must cover every participant: simulated users made 40 of the cohort's 311 calls, and a matcher that took calls from assistant messages only flagged all 40 of their results as unmatched.

\textbf{Test 3: continue after a failure.}
Make a required check fail while packets from the failed or an earlier run remain, then invoke the normal downstream entry point.
Require that no output of the failed run, or stale output substituted for it, reaches the judge or recording stub.
As in build failure handling~\cite{gnumake2023}, failed prerequisites must block dependent work.
A manifest written by a failed build can match its defective output exactly, so a matching hash does not authorise use.

\textbf{Test 4: change a checked file before use.}
For file-based inputs, change a staged packet after staging and, separately, after the last check before dispatch.
Require refusal before use in both cases.
P2's hash comparisons from staging to use resemble the file-integrity checks of the in-toto framework~\cite{torresarias2019intoto}.
Exact-byte hashes recorded at build and checked at staging would also detect changes made between build and staging, including the line-ending change.
J0's file-read tool returned the packet without checking its hash.
For such a judge, verification at read time or delivery of the validated bytes without another unchecked read can close the gap.
For other delivery modes, test the boundary between the last check and use.

\textbf{Scope.}
This exploratory case study examines one pipeline selected after its rendering omission was found, with 30 trajectories from 22 distinct user requests.
The controls were developed on the same cohort, tested on two cases without injected faults and evaluated on constructed fault cases written with knowledge of the code.

\textbf{Future evaluation.}
We plan to test transfer on two targets: AgentRewardBench's trajectory-to-judge pipeline~\cite{lu2025agentrewardbench}, whose screenshot and accessibility-tree inputs require adapting test~1, and independently written judge adapters for held-out AgentProcessBench trajectories.
For each target, an assessor who has not seen its checkers will derive the input contract, applicable obligations and valid and faulty cases from its documentation and evaluation goal.
We will adapt the tests before seeing those cases.
Without seeing the test results, the assessor will classify observed pipeline behaviour against the required behaviour for each applicable obligation.
These classifications will provide the reference for counting detected violations and false alarms, both with denominators.
We will also record adaptation effort and whether repairs change judge results under the same configurations.
On the predefined cases, a test transfers if it detects every confirmed violation of its obligation without flagging compliant behaviour.
Where no case or no violation is confirmed for an applicable obligation, detection remains unassessed.

\section{Conclusion}\label{sec:concl}

LLM-based evaluation depends not only on the judge but also on its input.
In this case, omitted tool calls persisted despite matching hashes, and restoring calls throughout the packets changed J0's later-context contrast from 13 to 2.
Both checkers accepted an unmatched tool result in a mutation case, while constructed integration tests exposed failure-handling and check-to-use gaps.
These findings motivate four input-assurance obligations, each with a direct test.

\textbf{Artifact.}
The artifact repository provides the study plans, released observations, source code and scripts for reconstruction, replay and reanalysis: \url{https://anonymous.4open.science/r/evaluation-artifact-342B/}.


\bibliographystyle{IEEEtran}
\bibliography{refs}
\end{document}